# Research and teaching in physics at the University of Franche-Comté 1845-1970

John M. Dudley,[1] Jeanne Magnin,[1] Luc Froehly,[1] Jerome Salvi,[1] Pierre Verschueren,[2] Maxime Jacquot[1]

1. Université de Franche-Comté, Institut FEMTO-ST, CNRS UMR 6174, Besançon, France

2. Université de Franche-Comté, Centre Lucien Febvre, EA 2273, Besançon, France

Corresponding author: john.dudley@univ-fcomte.fr

**Abstract**

Recently uncovered archives at the University of Franche-Comté in Besançon (France) reveal a rich history of research and teaching in physics since the Faculty of Science was first established in 1845. Here, we describe a selection of notable activities conducted by the named Chairs of Physics during the period 1845-1970. We uncover a long tradition of major contributions to physics education and research, including the production of highly regarded physics textbooks that were widely used in Europe, as well as pioneering contributions to electron diffraction and microscopy, Fourier optics, and holography. These discoveries yield valuable insights into the historical development of physics research in France, and show how even a small provincial university was able to stay up-to-date with international developments across several areas of physics.

# 1. INTRODUCTION

The University of Franche-Comté was founded in the French town of Dole by Philippe Le Bon in 1423, before being transferred to Besançon in 1691 by Louis XIV. The university was the tenth founded in France (Université de Besançon 1933, Thiou 2018). Its initial academic structure was organized around law, theology and medicine, and it was not until much later, in 1845, that a Faculty of Sciences was definitively established. In the context of the university's 600$^{th}$ anniversary in 2023, staff were asked to search for archival materials in their possession in order to build a more complete history of academic development over the years. In fact, various studies concerning the history of optics at the university had been intermittently carried out since 2015, and a significant collection of instruments, documents and photographs had already been saved from destruction, although this was in a poor state and only partially catalogued. The anniversary celebrations provided the impetus to examine and catalogue this material in detail, and an immediate and intriguing discovery was a roll of photographic 35 mm negatives stored in a metal canister labelled *"Galerie des Ancêtres."* Upon examination, the roll was found to contain photographs of a framed collection of portraits of the university chairs of *general physics*, from the first chair Charles-Cléophas Person in 1845, to Jean-Charles Viénot who had taken up the chair in 1963. This is shown in Figure 1. Research later revealed that it was Viénot who had assembled these portraits to display during a conference held in Besançon in 1970 (see below).

This single roll of film was a key to unlock the history of physics at the university, as we were able to easily search within local and national archives for information about the careers of the successive chairs of physics. The aim of this paper is to present the results of this study covering the period of appointments from 1845 until 1963. In addition to the chairs shown in Figure 1, we also discuss two other prominent scientists who occupied the position of Academy rectors during this period. Our treatment is not exhaustive, but it is useful to gather in one place key information concerning the different physicists who made their marks in different ways. We place a particular emphasis on the chairs who worked in the field of optical physics. Indeed, although it was Pierre Michel Duffieux (in post at Besançon from 1945) who founded the first dedicated research laboratory in optics (as distinct from general physics), we show that there has been a strong local tradition in both research and teaching in optical physics dating from the very beginnings of the Faculty of Science. And even though the university was in a provincial setting, its faculty was well-aware of international developments across several areas of physics, and several became extremely well-known personalities in their fields.

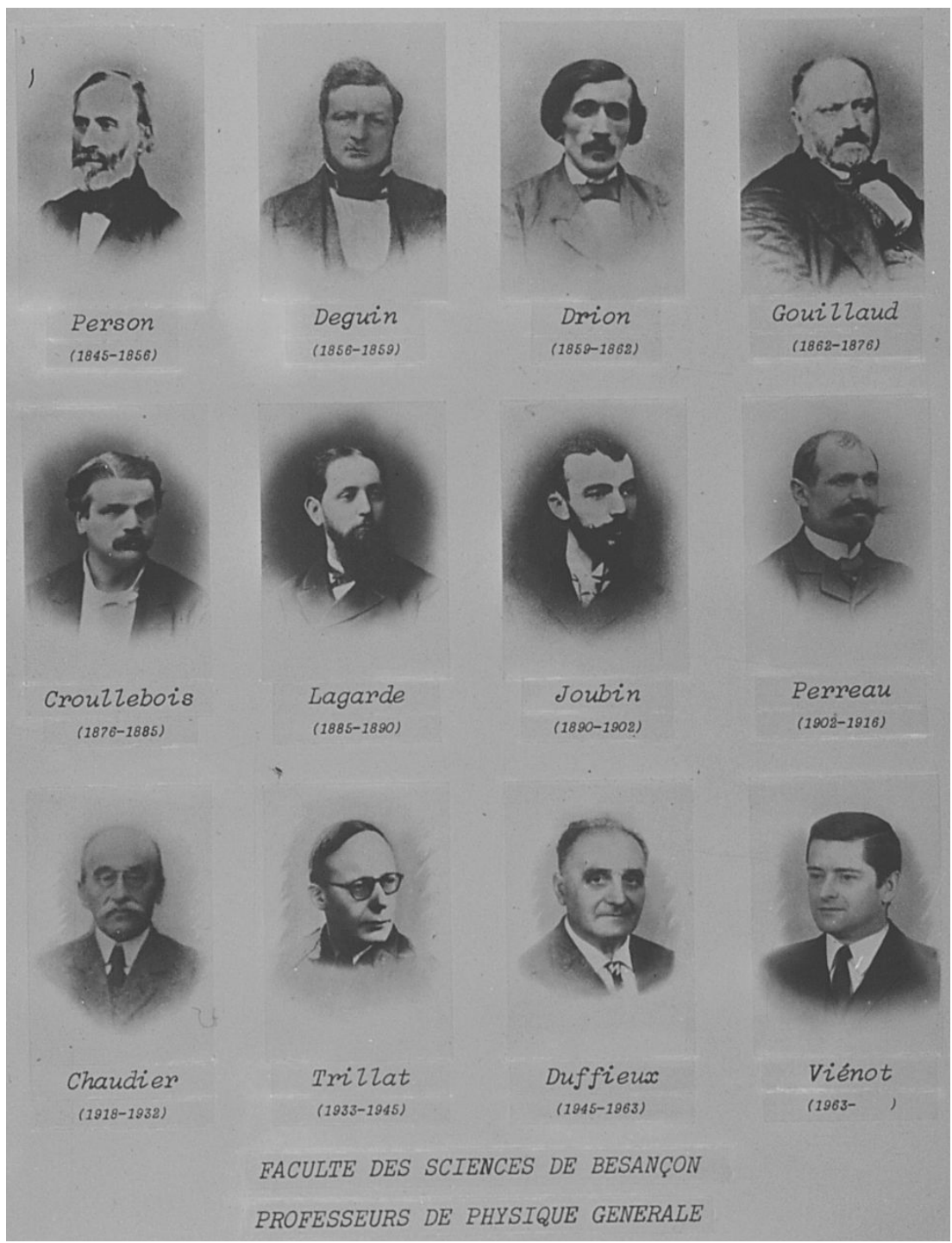

**Figure 1.** Professors of General Physics at the Faculty of Sciences of the University of Franche-Comté in Besançon from 1845 to 1963. The dates in each case seem to indicate the years of appointment to a post in Besançon, although this does not always correspond with the period where they officially had the title of Professor. This image is taken from a scanned 35 mm negative with sharpening and contrast enhancement.

## 2. THE CHAIRS OF PHYSICS.

**Charles-Cléophas Person** was the first chair of physics at the Faculty of Sciences in Besançon, nominated to the post in 1845. Person was born on 1 May 1801 in Mussy-sur-Seine (Aube), and died on 22 February 1884 in Nice (Alpes-Maritimes), aged 82. A more detailed biography appears in an obituary by (Count) Hilaire de Chardonnet (1839-1924) prepared for the Academy of Sciences of Besancon (de Chardonnet 1885). Person initially trained as medical doctor, but the majority of his career was in physics, and he obtained his doctorate (*Doctorat ès sciences physiques*) after examination defences on 7 and 9 August 1832 (Huguet and Noguès 2011). Obtaining a doctorate in France between 1815 and 1846 required the submission and defence of two dissertations (Maire 1892, Hulin 1990), and Person's chosen subjects were thermodynamics (physics) and atomic structure and chemical formulae (chemistry) (Person 1832a, 1832b). Also in 1832, Person ranked second in the national *agrégation* examination, which was (and is) a high-level selective teaching qualification (Cherval 2015). It was common in the 19$^{th}$ century for future university lecturers or chair professors to spend several years teaching at secondary level (even conducting research where possible) before a suitable university position became vacant. This was indeed the case for Person who taught at schools in Nancy, Metz, and Rouen. For further context into the structure of 19th century academia in France, see Shinn (1979), Nye (1986), and Fox (2012).

Person's appointment at Besançon in 1845 was made at the time when the Faculty of Science was established and administered under the deanship of the celebrated chemist (pioneer in the study of Aluminium) Henri Étienne Sainte-Claire Deville (1818-1881). Person would have had the responsibility of organizing courses in physics, likely benefiting from his previous teaching experience and from having written a textbook (Person 1836-1841). University archival records around 1844 suggest that Deville had been sent in advance to Besançon to organise logistics for the Faculty of Science while Person remained in Paris, selecting and purchasing experimental equipment with which to equip the new teaching laboratories (Archives Nationales F/17/14487). Indeed, the archives include an extensive inventory of optical, acoustical, and other instruments with a total cost of 22,000 Francs. The University of Franche-Comté has preserved a significant collection of these instruments, many of which remain in working condition today (Magnin et al. 2023). Person replaced Sainte-Claire Deville as dean of the Faculty in 1851 but, worrying about failing eyesight, he took early retirement in 1856. He is buried in the *Cimetière du Château* in Nice.

Person's research included publications on local meteorology, thermodynamics, and astronomy (Person 1847a, 1848, 1852; Wisniak 2017). In optics, he is especially recognized for his contribution to the understanding of the operation of the "Chinese Magic Mirror" (Person 1847b). Such mirrors are circular reflectors made of bronze whose polished front side reflects in the usual way to yield a virtual image. However, the back side has an embossed pattern, and the "magic" is associated with the fact that when the front surface is illuminated by a strong light source (e.g. sunlight), the pattern on the back can be projected to form a real image on a screen. Understanding the imaging properties of these mirrors fascinated researchers in Europe in the mid-19th century, and Person's 1847 letter to the French Academy of sciences was significant in pointing out the role of local surface deformations in the projection of the back-surface pattern. Person's priority in providing the first correct explanation was noted by Augustin Bertin (1818-1884)[1] in Bertin (1880), as well as in the 1911 Encyclopaedia Britannica (*Mirror* 1911). From a modern perspective, we now understand that the embossed back-side pattern is actually reproduced in sub-micron relief on the front surface, and the projected image arises from pre-focal ray deviation or "Laplacian" imaging (Mak and Yip 2001; Berry 2006). Interestingly, the imaging characteristics of magic mirrors have found recently been demonstrated experimentally using a flat optical window, namely a reconfigurable spatial light modulator that imposes an appropriate phase pattern on incident light (Hufnagel et al. 2022). This work has potential in both display and quantum information technologies.

**Nicolas Deguin** was born on 7 May 1809 in Autun (Saône-et-Loire), which is also where he died on 30 April 1860, aged 50 (Archives Départementales Saône-et-Loire; Archives Nationales Léonore). He entered the *Ecole préparatoire* (soon to be renamed *Ecole Normale Supérieure*) in 1828 and ranked third in the national *agrégation* examination in 1831. He taught at secondary level in Rodez and Toulouse, and obtained his doctorate (*Doctorat ès sciences physiques*) in 1832 for research in etherification (chemistry) and electromagnetism (physics) (Deguin 1832a, 1832b). His theses were defended on 25 and 31 July 1832 and, in what was quite an exception for the time, he presented before the Faculty of Science in Toulouse (rather than in Paris). He continued to teach at secondary level in Toulouse and Lyon until he was appointed professor of physics at the Faculty of Science of Clermont-Ferrand in 1854. In 1856 he replaced Charles-Cléophas Person as the chair of physics in Besançon, where he also took over the position of dean of the Faculty of Science the same year. He retired due to ill health in 1860.

Apart from his doctoral studies, Deguin does not appear to have published any research of note. Rather he appears to have been strongly focussed on writing textbooks for secondary school students. He had remarkable success (Khantine-Langlois 2018). For example, his *Cours élémentaire de physique* (Paris: Belin) was first published in 1836 and was republished multiple times before 1859. His *Cours élémentaire de chimie* (Paris: Belin) was similarly popular, published in several editions between 1845 and 1854. By the middle of the 19th century, his books were used in France, Belgium, and Switzerland (Bensaude-Vincent et al. 2002), selling over 30,000 copies (García-Belmar et al. 2005). And even in 1869 a decade after he died, his work (and name) evidently still had value, being adapted by the prolific Emile Gripon (1825-1912) in his *Precis de physique de Deguin* (Paris: Belin).

---

1. Note that Auguste Bertin is also known as Pierre Augustin Bertin-Mourot; see Archives Nationales (Pierrefitte-sur-Seine, France) F17/20159.

The success of Deguin's textbooks was widely known at the time, and indeed the favourable report on his candidature at the Faculty of Sciences of Clermont-Ferrand mentions this explicitly (Bensaude-Vincent et al. 2002). Deguin is typical of many textbook authors of the 19[th] century in that, even though his name would have been familiar to generations of students at the time, his contributions to science education are now mostly forgotten. Interestingly, it has been pointed out that the publication of Deguin's books across so many years of the 19[th] century reveals how innovations in printing had major influence on textbook formatting (Khantine-Langlois 2018). Specifically, early editions of Deguin's books have illustrations (based on copper engravings) in the form of separate plates collated at the end of the book, whereas later editions have the illustration (based on wood engravings) interspersed throughout the text (Bensaude-Vincent et al. 2002). In fact, the study of French scientific textbooks of this period is an active area of research, and their rich illustrations are historically important because they allow a modern reader to understand how 19[th] century physics and chemistry instruments were used and for what purpose (Khantine-Langlois 2005; Gires 2016). This has certainly been the case with the archival instrument collection in Besançon (Magnin 2023).

**Charles Alexandre Drion** was born on 10 July 1827 in Saverne (Bas-Rhin), and died on 3 April 1862 in Strasbourg (Bas-Rhin), aged 35 (AD Bas-Rhin). He entered the *Ecole Normale Supérieure* in 1847 and was ranked first in the national *agrégation* examination in 1850. He taught for short periods in secondary schools in Tournon, Strasbourg and Orléans, and then from 1853-1859 in Paris and Versailles. He defended his doctoral thesis (*Doctorat ès sciences physiques*) on 16 May 1859 (Drion 1859). At the Faculty of Sciences of Besançon, he was appointed *chargé de cours* in 1860 and professor of physics in 1862, the year he died. Like Deguin, Drion's career was defined by the writing of a successful secondary school textbook. The textbook was co-authored with Émile Jacques Fernet (1829-1905), with the first edition of Ch. Drion and E. Fernet, *Traite de Physique Elementaire* (Paris: Masson) published in 1861. It was a substantial book and richly illustrated. Although Drion did not live to see the fruits of his labour, the textbook was extremely successful and was republished in 12 editions up until 1893 (entry for *Fernet* in Havelange et al. 1986).

**Hippolyte Joseph Gouillaud** was born on 11 December 1816 in Buvilly (Jura), and died on 26 February 1877 in Besançon (Doubs), aged 60 (Archives Nationales Léonore). Although we not do know the full detail of his early studies, he began his teaching career in 1840, and during the following years taught at schools in Bourges, Grenoble, and Moulins (Corriger 2019). He was ranked fourth in the national *agrégation* examination in 1847, and was appointed to teach at the *lycée* in Besançon in 1853. In 1854 he obtained his doctorate (*Doctorat ès sciences physiques*) with a dissertation on heat conductivity in metals (Gouillaud 1854). In 1863 he was appointed *chargé de cours* at the Faculty of

Science in Besançon, and in 1864 he was nominated as professor (Corriger 2019). As remarked upon in a short obituary published in the journal of a local learned society (*Société d'émulation du Doubs*), he had been afflicted by severe gout since his childhood which impacted on his quality of life ("*il ne connut que l'austère jouissance du travail*" (Saillard 1878). Despite this difficulty, in addition to his thesis, he also published an article on magnetic properties of iron in the journal of the *Société d'émulation du Doubs* in 1863 (Gouillaud 1864). He retired in December 1876 and died two months afterwards.

**Désiré Marcel Croullebois** was born on 24 March 1843 in Bleury (Eure-et-Loire), and died on 19 May 1886 , aged 43 in Besançon (Doubs)[2]. He entered the *Ecole Normale Supérieure* in 1864, was ranked first in the national *agrégation* exam in 1867, and obtained his doctorate (*Doctorat ès sciences physiques*) after an oral defence on 19 April 1869 (Croullebois 1869). He occupied positions at the Faculties of Sciences in Marseille and Poitiers before being appointed to the Faculty of Sciences in Besançon, first as *chargé de cours* in 1877 and then as professor in 1879 (Huguet and Noguès 2011). He was an active and productive researcher, publishing extensively on optical physics between 1870 and 1882 in the *Annales de chimie et de physique* and the *Comptes rendus hebdomadaires des séances de l'Académie des sciences*[3]. He also participated in the activities of the *Société d'émulation du Doubs*. His research included geometrical optics, polarisation and birefringence, measurement of refractive indices, and interference and diffraction. He seems to have understood the need for an international perspective in science, and analysis of scientific travel archives by Klimentchenko et al. (2009) reveal that while at Marseille, he had travelled to England to study advances in experimental optics and teaching methods, and to Switzerland, Russia, and Germany to visit various physics laboratories[4]. And while in Besançon, he co-authored a translation from English into French of the classic textbook by Jenkin (1885) on electricity and magnetism[5].

Croullebois also published an 1882 textbook on the optics of the thick lens, based on lectures he delivered in Besançon (Croullebois 1882). The preface describes his aim to avoid difficult calculations (associated particularly with the mathematical approach of Gauss (Gauss 1841) and rather to present a simpler geometrical treatment. He is clearly satisfied with what he has written: *"Je pense donc avoir rendu un véritable service à l'enseignement… la théorie de Gauss devient maintenant accessible à tous les élèves des classes de Mathématiques dans nos lycées."* However, Croullebois was not the first to develop such a geometrical description, and he was well aware of this. In an 1880 article extending Gauss's methods to reflecting telescopes (Croullebois 1880), he cites an earlier study of thick lenses by Bertin (1878), remarking that Bertin himself was inspired by Carl Neumann (1832-1925) who had developed a geometric method even earlier (Neumann 1866). Perhaps what Croullebois was claiming was that his was the first textbook on the topic suitable for the level of teaching in French schools. But in any case, by whatever route the geometrical analysis of lens optics eventually entered into the mainstream of teaching, the constructions used by Croullebois (see Figure 2) would be recognizable to any student of the subject today.

---

2. The date of death is taken from his headstone. He is buried in the *Cimetière de Saint-Ferjeux* in Besançon. Note that many documents invert the order of his given names and write Marcel Désiré Croullebois.
3. The *Annales de chimie et de physique* (Paris: Masson) and the *Comptes rendus hebdomadaires des séances de l'Académie des sciences* (Académie des sciences), Paris are both online at the Bibliothèque Nationale de France (gallica.bnf.fr).
4. This reference remarks that the career files for Croullebois are found under Archives Nationales F/17/20502 and F/17/22811.
5. The translator's preface notes that Croullebois fell seriously ill when the translation was nearing completion. Fleeming Jenkin (1833-1885) was a polymath Scottish physicist and engineer.

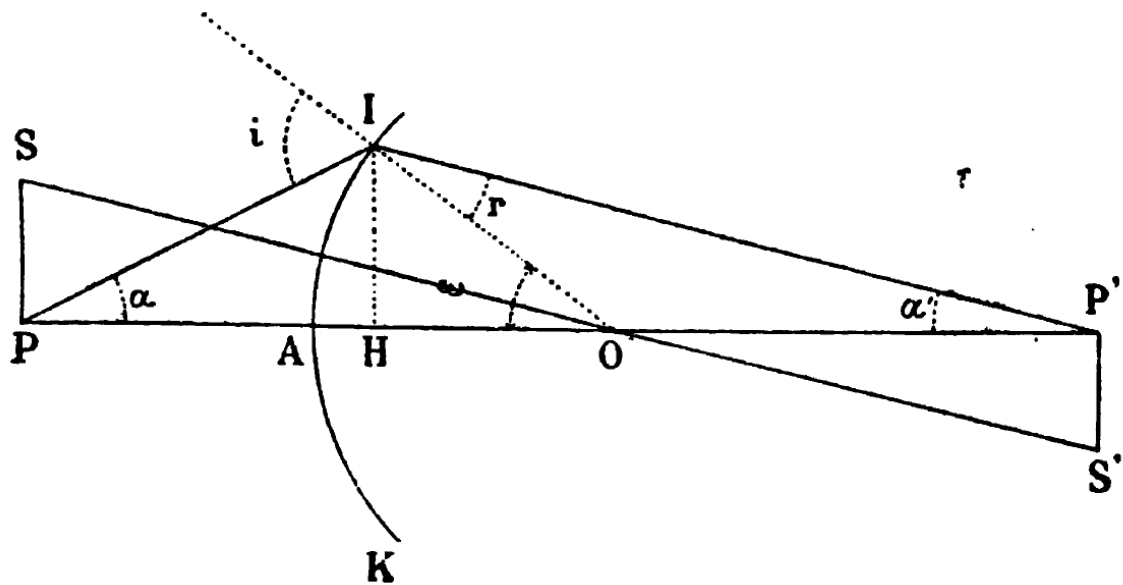

**Figure 2.** An illustration from the textbook on thick lens optics by Croullebois (1882), illustrating his use of geometrical constructions that would be familiar to any modern optics student.

**Charles Jacques Henri Lagarde** was born in Béziers (Hérault) on 27 October 1856, and died aged 33 on 20 April 1890 in Besançon (Doubs)[6]. A detailed account of his career appears in a published report of his eulogy delivered by the dean of the Faculty of Science at Besançon, Alexandre Vézian (1821-1903) (Vézian 1891). Henri Lagarde began his career in secondary education in 1876, teaching in Paris, then Montpellier and Bagnols (Gard). In 1878, he resumed his studies, as he obtained a scholarship at the Montpellier Faculty of Science. In 1881, he was ranked seventh in the national *agrégation* examination, and in 1882 he was appointed *maître de conferences* at Montpellier, although from 1883-1885, he taught at the Faculty of Besançon, replacing Croullebois who was on leave. After oral defence on 18 December 1884, Lagarde obtained his doctorate (*Doctorat ès sciences physiques*)

---

6. The date of death of the 20 April is deduced from a literal reading of his death certificate, notwithstanding that some published sources give the 21 April. He is buried in the *Cimetière des Chaprais* in Besançon, although his headstone is completely overgrown.

(Lagarde 1884) and in 1886 he was officially appointed as professor in Besançon. His doctoral thesis was not his first publication. As he worked his way upwards through the academic system from 1881-1886, Lagarde had published regularly in the *Compte Rendus de l'Académie des Sciences*, the *Annales de chimie et de physique*, and the *Bulletin de la Société Minéralogique*. His doctoral thesis describes a comprehensive experimental study of the spectral lines of hydrogen, and the 1885 version republished in *Annales de chimie et de physique* consists of a total of 123 pages with multiple detailed tables and figures (Lagarde 1885). This research was carried out over a 2-year period in Montpellier in the laboratory of André Crova (1833-1907), using a high resolution spectrometer that Crova himself had designed. To give a flavour of the laboratory environment of the time, Figure 3 shows the engraved illustration of Lagarde's experimental setup, clearly showing the spectrometer centrally positioned to analyse the emission from the discharge tube. The scientific instrument collection at the Université of Franche-Comté includes a Crova spectrometer, but we do not know for certain whether it was Lagarde himself who was responsible for its purchase.

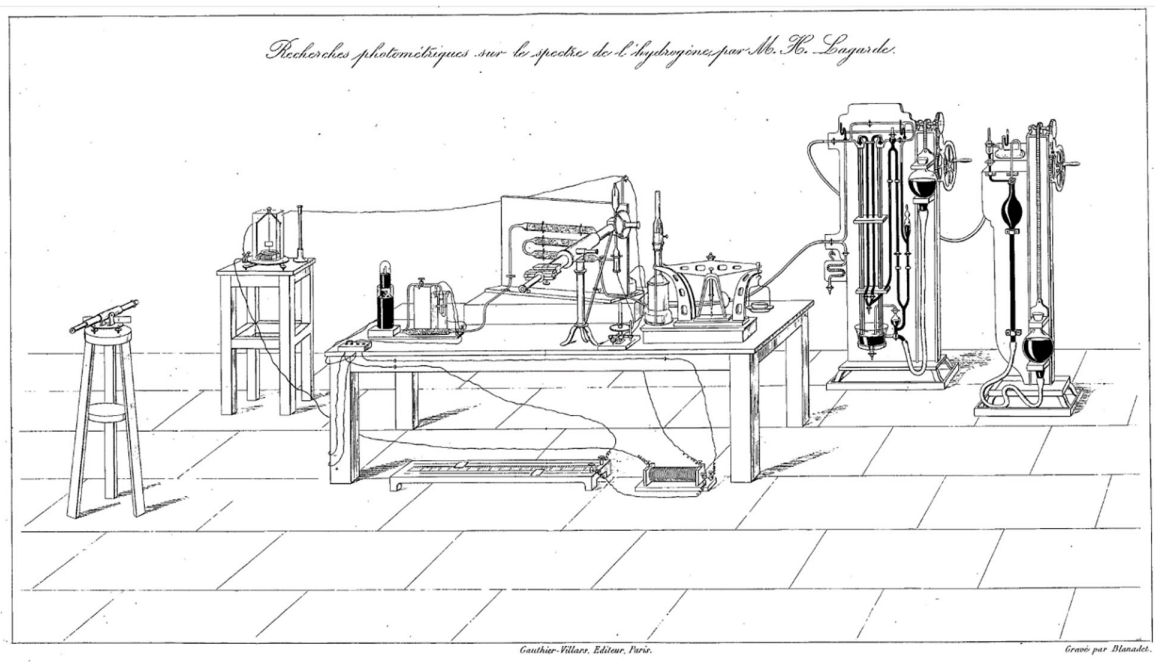

**Figure 3.** Lagarde's experimental setup for his doctoral research as illustrated in Lagarde (1885). The illustration clearly shows Crova's spectrometer centrally positioned on the laboratory bench to analyse the emission from the discharge tube. The illustration also shows the electrical system for producing the discharge high voltage, the system for generating the hydrogen gas by electrolysis, and the mercury pump system for generating a vacuum.

Vézian remarks in his eulogy that Lagarde's career successes at such a young age were achieved despite the disadvantage associated with not passing through the *Ecole Normale Supérieure* system (a comment that remains true to some extent even today!) Vézian also describes how Lagarde's commitment to all aspects of his academic career (teaching, research, administration) took its toll on his health, and that this became apparent even shortly after his nomination in 1886. But Vézian seems to have tried to make his appreciation of Lagarde's work known, remarking explicitly in the 1887-1888 report of the Faculty of Science in Besançon that Lagarde had relaunched an earlier tradition of weekly public experimental lectures (Vézian 1888). Vézian's eulogy describes how the death of Lagarde's son in 1889 took a further toll, and how in January 1890 he fell ill from an epidemic which was circulating in Besançon, possibly the Russian influenza of 1889-1890 which reached its peak in Besançon in January 1890 (Turquan 1893).

**Paul Jules Marie Joseph Joubin** was born in Angers (Maine-et-Loire) on 19 March 1862, and died in 1941 aged around 79 (Condette 2006). He entered the *Ecole Normale Supérieure* in 1882, was ranked seventh in the national *agrégation* exam in 1885, and obtained his doctorate (*Doctorat ès sciences physiques*) after oral defence on 20 June 1888. (Joubin 1888, 1889). His thesis studied the Faraday effect (rotation of the plane of polarised light by a magnetic field in certain materials), with the research carried out at the *École Polytechnique* in the laboratories of Alfred Cornu (1841-1902) and Alfred Potier (1840-1905). Joubin was *chargé de cours* in Montpellier in 1888 before occupying the same position at the Faculty of Science in Besançon in 1890. He was nominated as professor in Besançon in 1892, and was elected dean of the Faculty of Science in 1900. He does not appear to have been very active in research during his time in Besançon, possibly due to the sudden increase in physics teaching loads and student numbers associated with the 1894 creation of the *PCN* (*Certificat d'études physiques, chimiques et naturelles*), a qualification that was required for enrolment in medical school. This said, in 1896-1897, he did publish two linked papers using dimensional analysis to study analogies between mechanical and electromagnetic quantities (Joubin 1896, 1897). Joubin left Besançon in 1902 to continue his career in education administration: as rector of Chambery, Grenoble, and Lyon, before being appointed *Directeur de l'Instruction Publique* in French Indochina in 1922. He retired in 1925.

**François Perreau** was born in Cosne (Nièvre) on 24 July 1868, and died in Besancon (Doubs) on 30 June 1916, aged 47. His career is described in a published report of his eulogy, which was delivered by the highly distinguished mathematician Henri Padé (1863-1953), who was rector of the Academy of Besançon (Padé 1916). Perreau entered the *Ecole Normale Supérieure* in 1888, was ranked third in the national *agrégation* exam in 1891, and obtained his doctorate (*Doctorat ès sciences physiques*) in 1895 for a thesis prepared in the laboratory of Eleuthère Mascart (1837-1908) at the *Collège de France* (Perreau 1895). After an initial post as *maître de conferences* in Nancy, he was appointed professor in Besançon from 1902. He was dean of the Faculty of Science from February 1911, and appears to have been an effective organiser and administrator who was appreciated by his colleagues. Perreau's doctoral thesis studied the refractive and dispersive properties of gases, but his subsequent research activity seems to have been penalized by his commitment to teaching and administration. For example, amongst other activities, he created a new course on industrial electricity, funded by the municipality of Besançon and the Doubs *département*. Nonetheless, he did publish some articles related to electricity and its applications (Bolmont 2019), and he also had an editorial collaboration with the *Journal de Physique Théorique et Appliquée* providing commentaries on works from the English *Philosophical Magazine*[7]. Alongside his university tasks, he was actively involved in local affairs in Besançon. He played an important role improving public lighting in the town, and initiated a range of scientific and cultural activities aimed at the general public.

**Jules Chaudier** was born in Vauvert (Gard) on 6 May 1865 and died in Besancon (Doubs) on 28 February 1932, aged 66. His academic path included undergraduate (*Licence*) studies in both law and science; no records of him entering the *Ecole Normale Supérieure* or passing the *agrégation* exam have been located. He obtained a doctorate in law (*Doctorat en droit*) in December 1898 (Chaudier 1898) and then a doctorate in science (*Doctorat ès sciences physiques*) after oral defence on 8 May 1908 (Chaudier 1908a). He taught at the Faculty of Sciences in Montpellier, and was a member of the *Académie des sciences et lettres de Montpellier* from 1904 to 1910. After leaving Montpellier, he taught as *maître de conférences* at Grenoble before being appointed professor at the Faculty of Science of Besançon in 1918. He was appointed dean later in 1919. Jules Chaudier was still dean in 1931 when he died suddenly in the street. The circumstances of his death and its particular association with the cold weather was widely reported (Le Doyen 1932a, 1932b, Mort du doyen 1932). Aside from his thesis, Chaudier does not appear to have been especially active in research. Nonetheless, the publications associated with his thesis (Chaudier 1908b, Chaudier 1909) are interesting in that they describe the experimental setup in some detail, including how he used sunlight as the illuminating source through a diffusing screen after reflection from a heliostat (Gires 2016). The instrument collection at the University of Franche-Comté contains several models of heliostat.

**Jean-Jacques Marie Joseph Trillat** was born on 8 July 1899 in Paris, and died on 24 December 1987 in Versailles (Yvelines), aged 88. Trillat had a very distinguished career, eventually rising to be President of the French Academy of Sciences. We focus here on a selection of key events in his professional life, especially as they relate to his activities in Besançon (Trillat 1962, Takahashi 1988, Leprince-Ringuet 1988). After service in the First World War, Trillat studied at the *École municipale de physique et de chimie industrielles* in Paris from 1920 to 1923, graduating in the same year (39[th] *Promotion*) as future Nobel laureate Frédéric Joliot (1900-1958). Trillat notes in his memoire (Trillat 1962) that it was his father who encouraged him to follow up his engineering diploma with a doctorate, and he began doctoral studies at the Institut Pasteur in 1923. There is no record of Trillat sitting the *agrégation* examination. It was during the initial phase of his thesis that he met Maurice de Broglie (1875-1960) who reoriented Trillat's research around applying the new technique of X-ray diffraction to study the structure of long-chain organic compounds such as fatty acids, paraffins, and soaps. Trillat performed his experiments in de Broglie's private laboratory in rue Chateaubriand in Paris (Nye 1997), and obtained his doctorate (*Doctorat ès sciences physiques*) in 1926 (Trillat 1926a, 1926b). During this period, Trillat also met Maurice's brother Louis de Broglie, and thus became aware of the development of concepts such as wave particle duality and quantum mechanics. A photograph of Trillat with the de Broglies taken around 1925 is reproduced in Figure 4.

---

7. Some of his subsequent publications are authored under the name E. Perreau. The 1906 Census for Besançon lists his name as Francois Eugene Perreau.

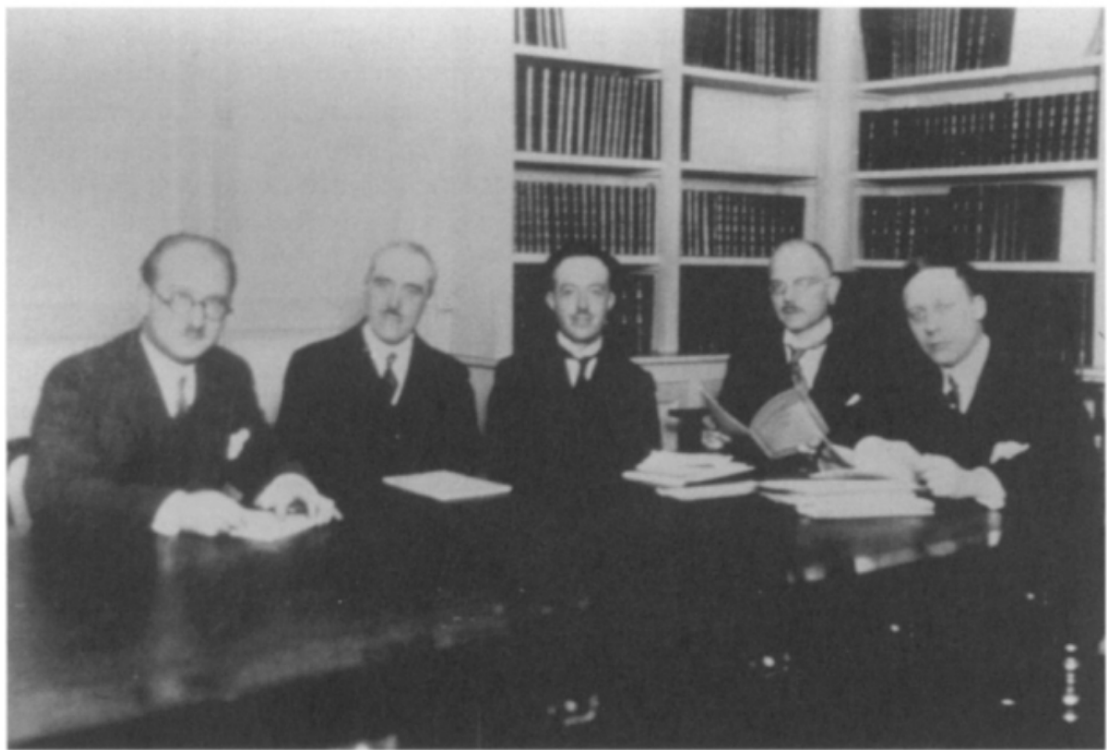

**Figure 4.** A photograph taken around 1925 in the library of Maurice de Broglie in his residence at rue Chateaubriand in Paris. Those seated are (left to right) Jean Thibaud, Maurice de Broglie, Louis de Broglie, Alexandre Dauvillier, Jean-Jacques Trillat. (Archives de l'Académie des Sciences, Paris. Biographical File of Louis de Broglie.)

After obtaining his doctorate, Trillat remained working with the de Broglies until 1933, extending his work in X-ray diffraction, and also developing expertise in electron diffraction following the 1927 experiments of Davisson and Germer and G. P. Thomson. In 1933, Trillat took up the chair in general physics in Besançon. Although the facilities upon arrival were very limited, he was able to bring with him some instruments that he had developed in de Broglie's laboratory. Trillat was clearly energetic in developing activities in Besançon, and by 1935 he was able to describe his progress in an article aimed at the French scientific community (Trillat 1935). He writes how his laboratory was now equipped with a range of X-ray diffraction sources and spectrometers, and the only two electron diffraction apparatus in France. He openly admits that the aim of his article is to make people aware of the existence of his laboratory, and to attract doctoral students to join him. Trillat's message here resonates still today – provincial French scientific laboratories continue to struggle for visibility in the shadow of the laboratories and universities in the much larger metropolitan centres of the country.

Trillat published extensively throughout his career (Notice Trillat 1959/1979). While in Besançon, he published both in the usual French journals of the time (*Comptes rendus hebdomadaires des séances de l'Académie des sciences, Journal de physique et le radium*), and in international journals (Transactions of the Faraday Society; *Annalen der Physik; Naturwissenschaften*.) In 1934, he authored a short monograph (in French) reviewing experimental evidence for wave mechanics (Trillat 1934). This work was introduced by a preface from Maurice and Louis de Broglie. Another paper of interest was a 1939 description of the use of shadow projection of a rotating reciprocal lattice to simulate the observation of an electron diffraction pattern (Trillat 1939). Trillat considered science from a broad perspective. In 1941 he published on the relation between basic research and industry (Trillat 1941), and later in 1970, he published an essay on the use of physics to give new insights into art (Trillat 1970).

In 1962, Trillat described how he had constructed France's first electron microscope in Besançon in 1935, together with his assistant Remi Fritz (1902-2002) (Trillat 1962). In a later memoire (Trillat 1981), Trillat described the device as "rudimentary", but it was nonetheless able to yield enlarged images of an oxide cathode (Grivet and Hawkes 2021). Unfortunately, Trillat himself never published anything on the device at the time, and the only contemporary description is in an article by Fritz (which unfortunately does not include any micrographs produced by the instrument) (Fritz 1936). Nevertheless, a photograph of the apparatus (Figure 5) was included as an accompanying figure (plate 2.1c) to Trillat (1981), and reproduced in Hawkes (2013). The importance of this early contribution to electron microscopy is widely recognized (de Gramont 1961, Hawkes 2013, van Gorkom et al. 2018).

Work in Trillat's laboratory ceased at the outbreak of the Second World War in 1939 (Trillat 1981), and regrettably all his equipment was lost in a road accident whilst it was being evacuated by truck to Switzerland in 1940 (Théobald 2011). Trillat applied his scientific expertise to the war effort in 1939 and 1940, initially developing inspection methods for shells and armour, and then assisting with the storage (hiding) of France's stock of radium and heavy water (180 kg obtained from Norway) (Leprince-Ringuet 1988). After the war, he moved to Paris to direct a research laboratory in electron diffraction and applications set up by the *Centre National de la Recherche Scientifique* (CNRS), and was later appointed at the Sorbonne. He was elected to the *Académie des Sciences* in 1959, and served as its President in 1974, during which time he was also President of the *Institut de France.*

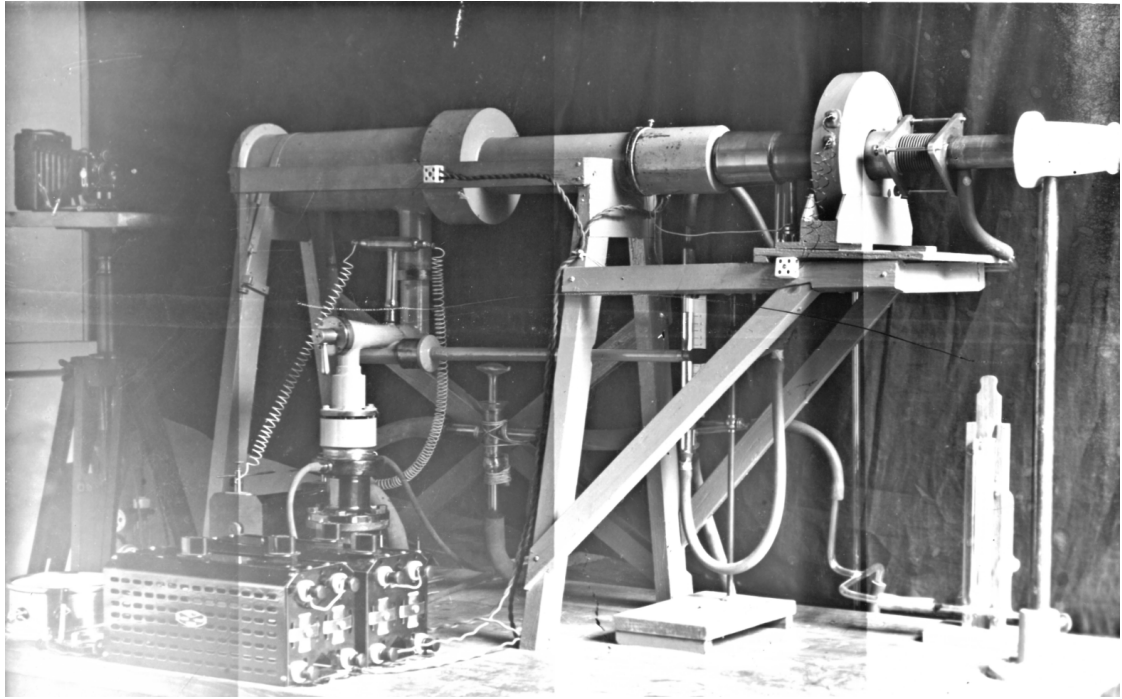

**Figure 5.** Photograph from around 1936 described as the first French electron microscope developed in Trillat's laboratory in Besançon (Archives de l'Académie des Sciences, Paris. Biographical File of Jean-Jacques Trillat.)

The University of Franche-Comté archives possess a number of physical items from the time of Trillat. In particular, the collection contains 3 Coolidge tubes dating from the 1930s (*Compagnie générale de radiologie*, Paris), as well as a 1931 X-ray spectrometer (Dr Müller's Improved X-ray Goniometer Spectrograph, made by Adam Hilger Ltd, London). There are also a number of glass photographic plates (negatives) showing results of various X-ray diffraction studies. The plates are in good condition in their original cardboard boxes and are of two sizes: 6 cm × 13 cm (18 boxes) and 9 cm × 12 cm (12 boxes). Most boxes and plates are undated, but the few dates that are written on the boxes or inscribed on the plates correspond to the period 1924-1931 i.e. predating Trillat's arrival in Besançon. One box even shows the name "Broglie" still visible despite a label pasted on top, and what can be read of the address is consistent with the second de Broglie laboratory in Paris at 12 rue Lord Byron. There are also some handwritten notes. Figure 6 shows a selection of this material. It is interesting to note the excellent state of preservation of these results after nearly a century; one wonders whether the digital data of today will be as accessible in the future.

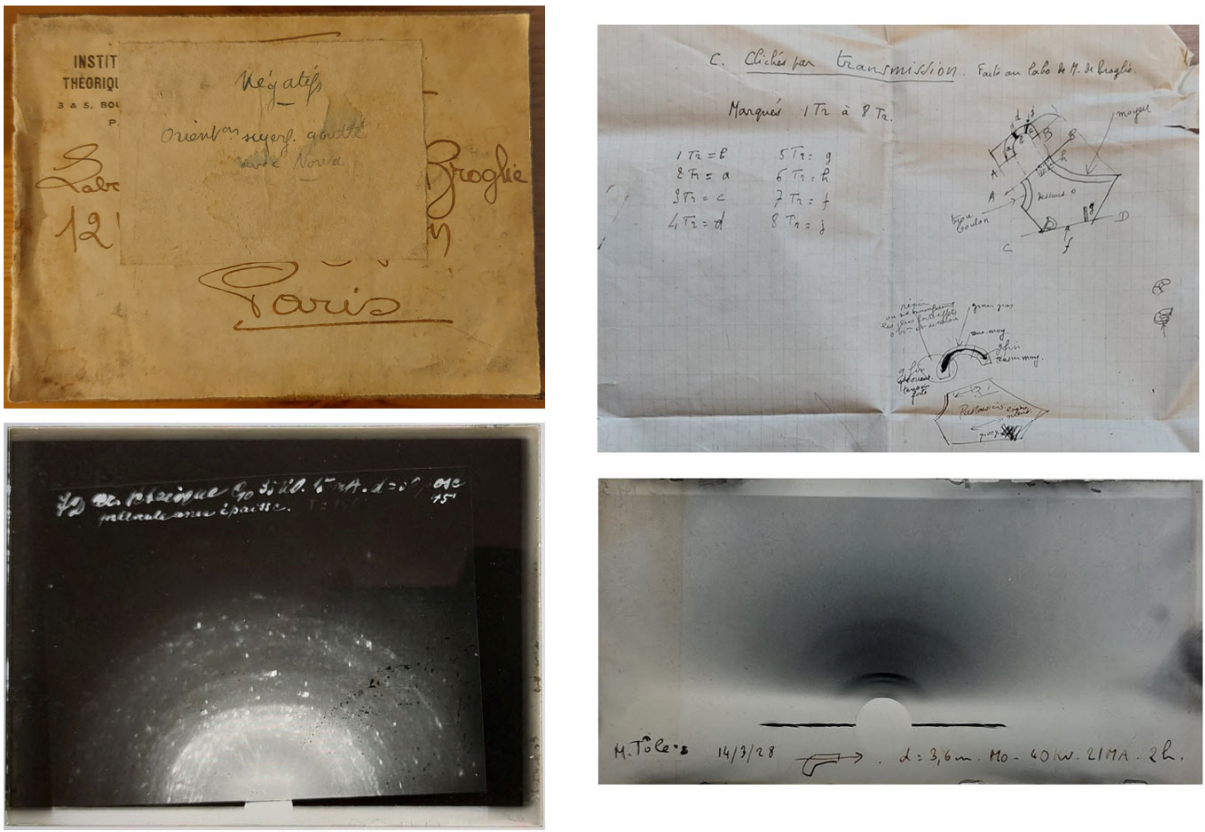

**Figure 6.** A selection of material from the photographic archives of Jean-Jacques Trillat held at the University of Franche-Comté. Top left: box containing 9 cm × 12 cm photographic plates, with the name and address of the de Broglie laboratory in Rue Byron partially visible. Bottom left: a plate from this box. Top right: some notes on the transmission geometry. Bottom right: Another example of a photographic plate, this time of size 6 cm × 13 cm. Credit: Archives of the Université de Franche-Comté.

**Pierre Michel Duffieux** was born on 21 February 1891 in Saint-Macaire (Gironde), and died on 3 June 1976 in Besançon (Doubs), aged 85. Two useful sources describing his career are a personal memoire (Duffieux 1974)[8], and a 1977 obituary by his colleague Pierre Mesnage (1910-2001) (Mesnage 1977). Duffieux entered the *Ecole Normale Supérieure* in 1912 but his studies were interrupted by the First World War. He served as second lieutenant from August 1914, but was evacuated sick a month later in September 1914, and discharged in September 1915. He voluntarily reenlisted in the Auxiliary Services in 1916, and from 1917 to 1919, he carried out scientific work in Bordeaux with Henri Bénard (1874-1939), studying refrigeration techniques for the storage and transport of meat. Duffieux notes in his memoire that it was during this period that he first learned about the Fourier transform, and also that he carried out some work in optics that had attracted the attention of Charles Fabry (1867-1945). Based on work performed during 1919 studying the thermal conductivity of insulating materials, Duffieux was later in 1921 to obtain his first publication (Duffieux 1921).

---

8. An earlier version of this text that had been written in 1970 was reproduced in Hawkes and Bonnet (1997).

After demobilisation in 1919, Duffieux returned to the *Ecole Normale Supérieure* and prepared for the *agrégation* examination which he passed (ranked third) in 1920. He moved to Marseille the same year to work with Fabry and, although Fabry relocated to Paris in 1921, Duffieux remained in Marseille and obtained his doctorate (*Doctorat ès sciences physiques*) after oral defence in June 1925 (Duffieux 1925, 1926). The members of his examination committee were Aimé Cotton (1869-1951), Fabry, and Robert Lespieau (1864-1947). Duffieux's thesis research involved the experimental spectroscopy of excited Hydrogen, Nitrogen, Cyanogen, and Carbon Monoxide, and his studies of Nitrogen in particular allowed him to participate in a lively international debate and publish in *Nature* in 1926.

A significant part of Duffieux's work used the Fabry-Perot interferometer, an instrument based on multiple beam interference developed around 1896 at Marseille by Fabry and Alfred Perot (1863-1925) (Fabry and Perot 1897, Connes 1986, Mulligan 1998). The Fabry-Perot interferometer has very high spectral resolution, and it rapidly found use in multiple areas, including pioneering work by Fabry with Henri Buisson (1873–1944) in the study of the Orion Nebula (1911), and in the discovery of the ozone layer (1913). It is worthwhile remarking here that in 1968, Duffieux lamented that Raymond Boulouch (1861-1937), his former teacher at the *lycée* in Bordeaux, was insufficiently recognized for his contribution to the theory of multiple beam interference (Duffieux 1968). Duffieux even went as far to propose that the "Fabry-Perot" should be renamed as the "Boulouch-Fabry," although this suggestion was not adopted (Connes 1986).

Duffieux left Marseille and took up a post in 1927 as *maître de conférences* at the Faculty of Sciences of Rennes. In 1934, Fabry asked Duffieux to look into a method of correcting for aberrations in the Fabry-Perot interferometer arising from residual non-uniformity of the plane reflecting surfaces. Duffieux worked on the problem and sent a solution, but Fabry didn't understand the mathematics, and bluntly stated that it would never be suitable for publication in the *Revue d'Optique*. Duffieux was discouraged, but an encounter with the mathematician Jean Dieudonné (1906-1992) reminded him that his analysis was in fact based on Fourier methods (that he had forgotten from his time in Bordeaux), and the purchase of a Mader-Ott harmonic analyzer (Hinzen et al. 2013) led him to the insight that a frequency-domain approach could yield an entirely new way to describe optical systems (Duffieux 1974). Duffieux developed his ideas more completely in a series of papers published between 1935 and 1940 (Duffieux 1935, 1939a, 1939b, 1939c, 1940a, 1940b) which together introduced the key concepts (spatial frequency, Fourier plane, convolution etc.) of what we describe today as Fourier optics. Duffieux also presented this work at the 1939 meeting of the French Physical Society but, as was the case with Fabry five years earlier, the audience did not understand his mathematical approach.

Discussions afterwards revealed that his colleagues were completely unfamiliar with the use of Fourier transforms, and so Pierre Fleury (1894-1976) suggested he write a book that included all the necessary mathematical explanations.

The first edition of *L'intégrale de Fourier et ses applications à l'optique* appeared as Duffieux (1946). By now, Duffieux had moved from Rennes to Besançon, appointed in 1945 as *maître de Conférences*, and then in 1949 as professor[9]. Unfortunately, because he only published in French (in an increasingly English-speaking scientific environment) Duffieux's book (and his many other papers) had comparatively little international visibility until 1959 when Born and Wolf explicitly cited his pioneering role in their *Principle of Optics* (Born and Wolf 1959). Pleasingly, there was also increasing recognition in France (Amat et al. 1962). Duffieux's book was reprinted in a corrected second edition in French in 1970 (Duffieux 1970), which was translated into Japanese in 1977 (Duffieux 1977), and finally into English in 1983 (Duffieux 1983). That an English translation was available was clearly to be welcomed, but at this stage it was largely of historical interest (Hawkes 1983, Harburn 1984). This is because Duffieux's approach to the subject was becoming outdated even at the time of the 1970 edition – Fourier techniques had become mainstream in physics since the 1950s (Lighthill 1958, Papoulis 1962) and an English-language textbook on Fourier optics using the latest language and terminology had appeared in 1968 (Goodman 1968). Goodman's book was to be become a classic.

Duffieux retired in 1963 after having spent the last years of his career designing a new and environmentally-stabilized optics laboratory for the university. The timing here was ideal because these new facilities allowed researchers in Besançon to take advantage of the invention of the laser in 1960 to become one of the first French laboratories to study lasers, coherent optics and the many associated applications. Around this time, a new "Laboratory of Optics" was created (administratively distinct from general physics), and this was headed by Duffieux's successor Jean-Charles Viénot who had been appointed Professor in his place in 1963. When Duffieux died in 1976, Viénot edited a special issue of *Optica Acta* in his memory which provides an important snapshot of the research areas being studied in Besançon during the 1970s (Special Issue 1977). An obituary written by André Maréchal (1916-2007) had also appeared earlier in the monthly magazine of the American Physical Society (Maréchal 1976). Duffieux is buried in the *Cimetière de Saint-Ferjeux* in Besançon.

**Jean-Charles Viénot** was born on 22 June 1930 in L'Isle-sur-le-Doubs (Doubs) and died on 26 September 2022. Viénot had been an undergraduate student in Besançon before doing doctoral studies at the University of London (Imperial College London) (Viénot 1958) and then returning to Besancon as *Maître de Conférences* in 1958. A particular highlight of Viénot's tenure was a large international conference on holography that took place in Besançon in July 1970. That such a conference was possible in Besançon was due both to the reputation of Duffieux as the founder of Fourier optics, as well as Viénot's international contacts through his position as Secretary-Treasurer of the International Commission for Optics. The conference was remarkable for its time, attracting over 400 attendees, including over 250 from outside France (Viénot et al. 1970, Viénot 1970). Figure 7 shows the conference photograph. Holography was in fact to become a major area of activity for the laboratory during the 1970s, with a particularly unique aspect of this work being collaboration with museums and artists (Tribillon and Fournier 1977, Bainier and Tribillon 1989). Viénot continued as director of the Laboratory of Optics until 1983 and remained in Besançon until 1987 when he moved to take up the direction of the *Institute Science Matière et Rayonnement* in Caen (Calvados).

---

9. Full details of Duffieux's career are to be found in: National Archives (Pierrefitte-sur-Seine, France) F/17/28127 and F/17/16784.

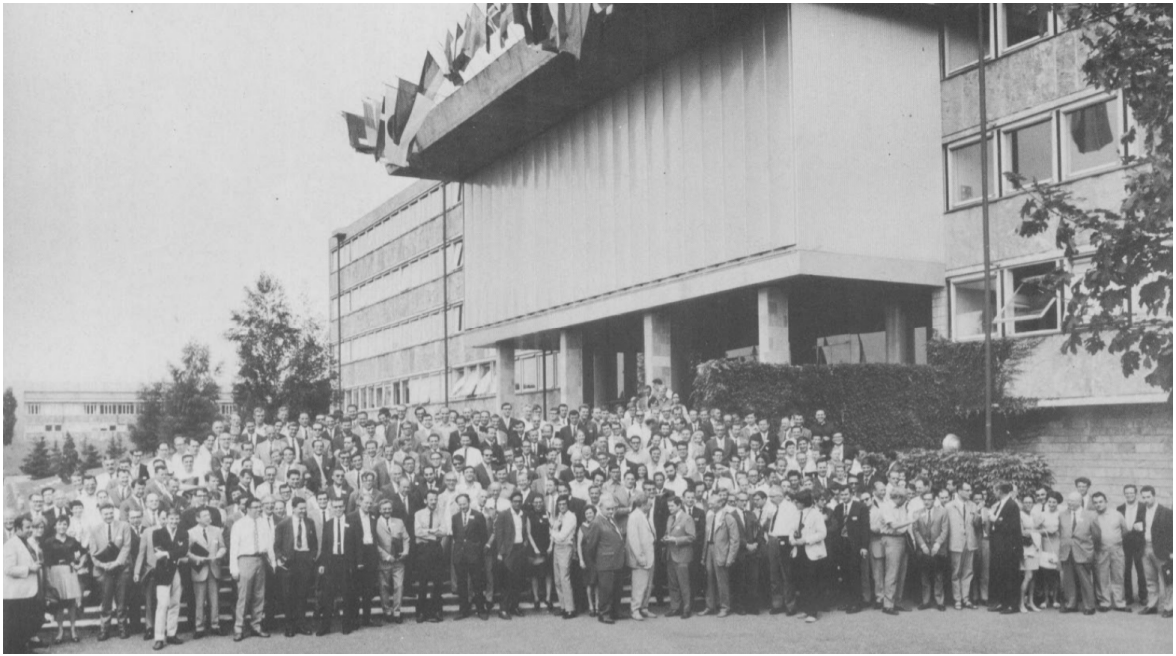

**Figure 7.** Conference photograph from the 1970 Applications of Holography conference, taken in front of the Faculty of Science in Besançon. Duffieux is the dark-suited man standing side-on in the front and centre of the photograph. Credit: Archives of the Université de Franche-Comté.

## 3. TWO PROMINENT RECTORS

In addition to the chairs of physics discussed above, a number of other prominent French scientists also spent parts of their careers in Besançon (Vernotte 2016)[10]. We briefly describe here two with strong connections to optics, both occupying the administrative role of rector of the Academy of Besançon. For background to the roles of rectors in French universities, see Condette (2009, 2016).

**Jean-Antoine Quet** was born 21 October 1810 in Nimes (Gard), and died on 27 November 1884 in Paris, aged 74. He was rector of the Academy of Besançon during the period 1854-1856 (entry for *Quet* in Havelange et al. 1986). He entered the *Ecole Normale Supérieure* in 1830 and ranked first in the *agrégation* examination in 1833. He taught at a secondary school in Grenoble in 1834, was *chargé de cours* at the Faculty of Science in Grenoble from 1834-1835, then spent his career from 1836 to 1854 teaching at secondary schools in Versailles and Paris. He obtained his doctorate (*Doctorat ès sciences mathématiques*) after oral defence on 2 July 1839 (Quet 1839). After leaving Besançon, he was rector of the Academy of Grenoble (1856-1864), before being appointed as Inspector General of Secondary Education until his retirement in 1883 (Notice Quet 1884). During both academic and administrative phases of his career, he regularly published research in diverse areas of physics (Notice Quet 1873), including a lengthy calculation on edge diffraction published while he was in Besançon (Quet 1856). This calculation does not appear to have been motivated by any practical consideration, and his aim rather appears to have been to confirm Fresnel's theory of diffraction for the case of a sequence of diffracting screens, a configuration that had not been considered before.

**Jules Antoine Lissajous** was born on 4 March 1822 in Versailles (Yvelines) and died on 24 June 1880 in Plombières-lès-Dijon (Côte-d'Or). He was rector of the Academy of Besançon during the period 1875-1879 (Condette 2006). He entered the *Ecole Normale Supérieure* in 1841 and then occupied a number of secondary teaching positions throughout France returning to Paris and teaching at the *Lycée Saint Louis* from 1849-1874. He ranked third in the national *agrégation* examination in 1847 and obtained his doctorate (*Doctorat ès sciences physiques*) after oral defence on 27 November 1850. (Lissajous 1850). The physics component of his thesis studied the nodal positions of vibrating metallic bars, and included use of the sand pattern method of Ernst Chladni (1756-1827). Lissajous continued to study wave properties after his thesis, and gained widespread attention for his 85-page paper in 1857 which described visualizing acoustic waves using successive reflections of light from orthogonal vibrating tuning forks (Lissajous 1857). This is a classic paper in the history of science, and *Lissajous Figures* are now taught as an essential part of introductory physics. Lissajous gave many well-received demonstrations of this technique, with John Tyndall (1820-1893) attending and recording one delivered at the Royal Institution of London in 1857 (Taylor 1988). Unfortunately, while Lissajous's research was held in very high regard, this seems to have been at the expense of his appointed position as a secondary school teacher. Regular reports by his superiors express concern about his lack of commitment to his duties and students, and his carelessness in classroom presentation (Condette 2006, Brasseur 2010). Later in his career he was transferred (at his request) to senior administrative roles, first as rector in Chambery in 1874, and then as rector in Besançon from 1875-1879. His activities in Besançon appear to have been mainly organizational, although he did participate in the activities of the *Societé d'émulation du Doubs.* He retired in 1879 and died in 1880. He is buried in Plombières-lès-Dijon.

---

10. See also the memoires of the *Société d'émulation du Doubs* or the *Académie des sciences, belles lettres et arts de Besançon et de Franche-Comté*.

## 4. CONCLUSIONS

The biographies above illustrate the varied contributions made by key personalities in physics in Besançon from 1845 to 1970. While the coverage cannot claim to be exhaustive in any way, our hope is that it will serve as a compilation of essential information to motivate further and more complete historical studies. Nonetheless, even as it stands, we clearly see a rich tradition in physics at the University of Franche-Comté that spans important thematic areas, particularly the field of optical physics. However, somewhat to our surprise, in preparing this article and speaking with over 40 retired members of staff from the University of Franche-Comté who were working in the 1950s and 1960s, only one was aware of any heritage prior to Pierre-Michel Duffieux. And even then, only the names of Trillat and Lissajous were remembered – the contributions of the others had been lost to obscurity.

On a general level, we believe that the loss of such local historical knowledge is a great loss to the collegial life and the role of a university. But perhaps even more importantly, it is also a missed opportunity from a pedagogical perspective. Teaching the historical development of a subject is an important part of university education, and being able to position pedagogy within a rich local landscape adds great value to classroom discussions. Of course, such a history is already preserved and celebrated in many large universities around the world, but it is likely that nearly all faculties of science with origins in the 19$^{th}$ century would be able to tell similar stories to that which we have presented here.

## 5. ACKNOWLEDGEMENTS

This work was supported by Agence Nationale de la Recherche (projects OPTIMAL ANR-20-CE30-0004 and HOLO-CONTROL ANR-21-CE42-0009) and by the French Investissements d'Avenir program, through the cross-disciplinary research (EIPHI) Graduate School (ANR-17-EURE-0002).

## 6. APPENDIX - ADDITIONAL SOURCES

For convenience, the references cited in this paper are primarily readily-available published papers, books, and databases. However, each of the scientists above is also the subject of an administrative file held in the French national archives (Pierrefitte-sur-Seine, *Archives Aationales* AN) and many also have another administrative file at the level of the local départment of the Doubs (Besançon, *archives départmentales* AD). These contain additional information of interest for further study, and for completeness we list them here as follows: Person AN F/17/21472 and AD T1458; Deguin AN F/17/20540 and AD T1397; Drion AN F/17/20630 and AD T1400; Gouillaud AN F/17/20862/B and AD T1415; Croullebois AN F/17/20502 and F/17/22811 (F/17/2951 for his mission to Russia) and AD T1394; Lagarde AN F/17/22934 and AD T1430. Joubin AN F/17/23773 and AD T1525; Perreau AN F/17/25879 and AD T1457; Chaudier AN F/17/26722 and AD T1501; Trillat AN F/17/30128 ; Duffieux AN F/17/28127 and F/17/16784; Quet AN F/17/21560 and AD T1463; Lissajous F/17/21188 and AD T1436. Further administrative information relating to appointments, retirements, student numbers etc. can be found in the series of annual reports: *Rapports sur la situation et les travaux des etablissements d'enseignement superieur de Besançon* published between 1897 and 1948.